\documentclass[conference]{IEEEtran}
\usepackage[shortlabels]{enumitem}
\usepackage[nospace,noadjust]{cite}
\usepackage{graphicx} 
\usepackage{epsfig} 
\usepackage{amsmath,amsthm,amssymb}  
\usepackage{float}
\usepackage{cases,setspace,adjustbox}
\usepackage{array}
\usepackage[update,prepend]{epstopdf}
\usepackage{eqparbox}
\usepackage{url}
\usepackage{caption, subcaption}
\usepackage{xcolor}
\usepackage{mathtools}
\usepackage{bbm}
\usepackage{multirow}
\usepackage{multicol}

\usepackage[acronym]{glossaries}
\usepackage{comment}
\usepackage{xcolor}
\newcommand*\circled[1]{\kern-2.5em%
  \put(0,4){\color{white}\circle*{11}}\put(0,4){\circle{11}}%
  \put(-3,0){\color{black}\large#1}~~}
\newcommand\blfootnote[1]{%
  \begingroup
  \renewcommand\thefootnote{}\footnote{#1}%
  \addtocounter{footnote}{-1}%
  \endgroup
}
\setacronymstyle{long-short}

\newacronym{AR}{AR}{Augmented Reality}
\newacronym{arima}{ARIMA}{Autoregressive Integrated Moving Average}

\newacronym{cdf}{CDF}{cumulative distribution function}
\newacronym{cnn}{CNN}{Convolutional Neural Network}
\newacronym{comp}{CoMP}{Coordinated Multi-Point}
\newacronym{convlstm}{ConvLSTM}{Convolutional LSTM}
\newacronym{cp}{CP}{Cyclic Prefix}
\newacronym{crs}{CRS}{Cell Specific
Reference Signal}
\newacronym{ctl}{CTL}{Communications Technology Laboratory}

\newacronym{d2d}{D2D}{Device-to-Device}
\newacronym{dc}{DC}{dual connectivity}
\newacronym{dci}{DCI}{Downlink Control Information}
\newacronym{dl}{DL}{downlink}
\newacronym{dsrc}{DSRC}{Dedicated short-range communications}
\newacronym{dss}{DSS}{Dynamic Spectrum Sharing}
\newacronym{ddpg}{DDPG}{Deep Deterministic Policy Gradient}
\newacronym{dmaap}{DMaaP}{Data Movement as a Platform}
\newacronym{ets}{ETS}{Exponential Smoothing}
\newacronym{embb}{eMBB}{enhanced mobile broadband}
\newacronym{enb}{eNB}{Evolved Node-B}
\newacronym{epc}{EPC}{Evolved Packet Core}
\newacronym{emtc}{eMTC}{enhanced MTC}

\newacronym{fdd}{FDD}{Frequency Domain Duplex}
\newacronym{fdm}{FDM}{Frequency Domain Multiplexing}
\newacronym{firstnet}{FirstNet}{First Responder Network Authority}

\newacronym{gan}{GAN}{Generative Adversarial Network}
\newacronym{gps}{GPS}{Global Positioning System}
\newacronym{gnb}{gNB}{Next Generation Node-B}

\newacronym{harq}{HARQ}{Hybrid Automatic Repeat reQuest}

\newacronym{iiot}{IIoT}{Industrial Internet of Things}
\newacronym{its}{ITS}{Intelligent Transport Systems}
\newacronym{itu}{ITU}{International Telecom Union}

\newacronym{LEO}{LEO}{Low Earth Orbit}
\newacronym{lte}{LTE}{Long Term Evolution}
\newacronym{lteapro}{LTE-A~Pro}{LTE Advanced Pro}
\newacronym{lstm}{LSTM}{Long Short-Term Memory}

\newacronym{mcs}{MCS}{Modulation and Coding Scheme}
\newacronym{mcptt}{MCPTT}{Mission-Critical Push-to-Talk}
\newacronym{mimo}{MIMO}{Multiple-Input Multiple-Output}
\newacronym{mmtc}{mMTC}{massive machine-type communication}
\newacronym{mtc}{MTC}{Machine-Type Communication}
\newacronym{mlp}{MLP}{Multilayer Perceptron}
\newacronym{ma}{MA}{Moving Average}
\newacronym{mm}{MM}{Moving Median}
\newacronym{mae}{MAE}{Mean Absolute Error}
\newacronym{mdp}{MDP}{Markov Decision Process}

\newacronym{nist}{NIST}{National Institute of Standards and Technology}
\newacronym{nbiot}{NB-IoT}{Narrowband-Internet of Things}
\newacronym{nr}{NR}{New Radio}
\newacronym{ntia}{NTIA}{National Telecommunications and Information Administration}

\newacronym{oran}{O-RAN}{Open Radio Access Network}
\newacronym{owl}{OWL}{Online Watcher of LTE}

\newacronym{prosas}{ProSAS}{Proactive Spectrum Adaptation Scheme}
\newacronym{prb}{PRBs}{Physical Resource Blocks}
\newacronym{psc}{PSC}{Public Safety Communications}
\newacronym{pscch}{PSCCH}{Physical Sidelink Control Channel}
\newacronym{psfch}{PSFCH}{Physical Sidelink Feedback Channel}
\newacronym{pssch}{PSSCH}{Physical Sidelink Shared Channel}

\newacronym{qos}{QoS}{Quality of Service}

\newacronym{ran}{RAN}{Radio Access Network}
\newacronym{rat}{RAT}{Radio Access Technology}
\newacronym{re}{RE}{Resource Element}
\newacronym{ric}{RIC}{RAN Intelligent Controller}
\newacronym{relu}{ReLU}{Rectified Linear Unit}
\newacronym{rmse}{RMSE}{root-mean-square error}
\newacronym{rnn}{RNN}{Recurrent Neural Network}
\newacronym{rl}{RL}{Reinforcement Learning}

\newacronym{sdr}{SDR}{Software Defined Radio}
\newacronym{sl}{SL}{sidelink}
\newacronym{smo}{SMO}{Service and Management Orchestration}

\newacronym{tdd}{TDD}{Time Domain Duplex}
\newacronym{tdm}{TDM}{Time Domain Multiplexing}
\newacronym{timegan}{TimeGAN}{Time-Series Generative Adversarial Network}
\newacronym{tr}{TR}{Technical Report}
\newacronym{td3}{TD3}{Twin Delayed Deep Deterministic Policy Gradient}
\newacronym{td}{TD}{temporal difference}
\newacronym{urllc}{URLLC}{ultra-reliable low-latency communication}
\newacronym{uav}{UAVs}{Unmanned Aerial Vehicles}
\newacronym{ue}{UE}{User Equipment}
\newacronym{ul}{UL}{uplink}

\newacronym{v2v}{V2V}{Vehicle-to-Vehicle}
\newacronym{v2x}{V2X}{Vehicle-to-Everything}
\newacronym{VR}{VR}{Vitual Reality}

\newacronym{wnd}{WND}{Wireless Networks Division}
\newacronym{wlan}{WLAN}{Wireless Local Area Network}
\newacronym{3gpp}{3GPP}{3rd Generation Partnership Project}
\begin{document}
\title{AdapShare: An RL-Based Dynamic Spectrum Sharing Solution for O-RAN}
\author{Sneihil Gopal$^{\ddagger\dagger}$, David Griffith$^{\star}$, Richard A. Rouil$^{\star}$ and Chunmei Liu$^{\star}$ \\
$^{\ddagger}$PREP Associate, National Institute of Standards and Technology (NIST), USA\\
$^{\dagger}$Department of Physics, Georgetown University, USA\\
$^{\star}$National Institute of Standards and Technology (NIST), USA.\\
Emails: \{sneihil.gopal, david.griffith, richard.rouil, chunmei.liu\}@nist.gov}
\maketitle
\section*{Abstract}
\label{sec:abstract}
The Open Radio Access Network (O-RAN) initiative, characterized by open interfaces and AI/ML-capable RAN Intelligent Controller (RIC), facilitates effective spectrum sharing among RANs. In this context, we introduce AdapShare, an O-RAN-compatible solution leveraging Reinforcement Learning (RL) for intent-based spectrum management, with the primary goal of minimizing resource surpluses or deficits in RANs. By employing RL agents, AdapShare intelligently learns network demand patterns and uses them to allocate resources. We demonstrate the efficacy of AdapShare in the spectrum sharing scenario between LTE and NR networks, incorporating real-world LTE resource usage data and synthetic NR usage data to demonstrate its practical use. We use the average surplus or deficit and Jain's fairness index to measure the RL system's performance in various scenarios. AdapShare outperforms a quasi-static resource allocation scheme based on long-term network demand statistics, particularly when available resources are scarce or exceed the aggregate demand from the networks. Lastly, we present a high-level O-RAN-compatible architecture using RL agents, which demonstrates the seamless integration of AdapShare into real-world deployment scenarios.

\section{Introduction}
\label{sec:intro} 
With the debut of 5G Advanced (5G Adv) and the preparations for 6G~\cite{ref:6G_saad2019_vision}, spectrum sharing will be crucial for efficient spectrum management. The requirements of 5G~Adv and the potential of 6G for higher data rates and a proliferation of connected devices underscore the need for optimal utilization of spectrum resources. Within the realm of 5G \gls{nr}, \gls{dss} emerges as a key radio technology enabling the simultaneous use of \gls{lte} and 5G within the same frequency band, i.e., the sub-6 GHz band from 410~MHz to 7.125~GHz. This innovative technology dynamically assesses real-time demand for both 5G and \gls{lte}, autonomously dividing the available bandwidth and dynamically allocating frequencies based on the needs of each mobile standard. \blfootnote{U.S. Government work, not subject to U.S. Copyright.}

In this paper, we propose a \gls{dss} scheme in the context of the \gls{oran} architecture~\cite{ref:oran_tr_wg1}. The \gls{oran} initiative~\cite{ref:oran_polese_2023}, with its open interfaces, virtualization, and \gls{ric} that supports artificial intelligence/machine learning (AI/ML) facilitates effective spectrum sharing in the \gls{ran}. Within this framework, we introduce AdapShare, an \gls{oran}-compatible \gls{dss} solution that leverages \gls{rl} for data-driven, intent-based spectrum management. 

AdapShare's objective is to minimize surpluses or deficits of spectrum resources; it employs \gls{rl} agents to learn radio resource demand patterns and intelligently allocate resources. We showcase the effectiveness of AdapShare in a spectrum sharing use case involving co-existing \gls{lte} and \gls{nr} networks. Our analysis uses real-world \gls{lte} resource usage data collected using BladeRF \gls{sdr}\footnote{Certain commercial equipment, instruments, or materials are identified in this paper in order to specify the experimental procedure adequately. Such identification is not intended to imply recommendation or endorsement by National Institute of Standards and Technology (NIST), nor is it intended to imply that the materials or equipment
identified are necessarily the best available for the purpose.\label{refnote}} and \gls{owl}, an open-source \gls{lte} sniffer~\cite{ref:owl_bui_2016}, as well as \gls{nr} data generated synthetically using a \gls{timegan}~\cite{ref:timegan}. To evaluate AdapShare's efficacy, we compare it with a quasi-static resource allocation scheme based on network demand statistics~\cite{ref:oran_gopal_2024prosas}. AdapShare outperforms the quasi-static scheme by dynamically adapting to real-time demand patterns, particularly when available resources are either scarce or when they exceed the
aggregate demand from the networks. 

Furthermore, AdapShare's \gls{oran} compatibility allows for flexible deployment options: as a \gls{ran} application (rApp) in the non-real-time (non-RT) \gls{ric} or as an external application (xApp) in the near-real-time (near-RT) \gls{ric}. In the first deployment scenario, an rApp executes \gls{rl} agents and communicates with an xApp for optimized resource allocation. In the second scenario, the xApp hosts \gls{rl} agents, observing the network state from the \gls{ric} and autonomously adjusting resource allocation. 

Additionally, we discuss a high-level \gls{oran}-compatible architecture to showcase AdapShare's seamless integration into real-world scenarios, emphasizing adherence to industry standards and reliance on real-world data. AdapShare is a promising approach to spectrum sharing that offers practical solutions for network operators to deploy advanced network architectures while ensuring optimal resource utilization and performance.

The structure of the paper is as follows. Section~\ref{sec:prior_work} delves into related work. Section~\ref{sec:proposed_framework} provides an overview of \gls{oran} and our proposed solution. Section~\ref{sec:methodology} outlines the methodology we used to develop and evaluate AdapShare. We demonstrate the effectiveness of AdapShare in Section~\ref{sec:results} and in Section~\ref{sec:integration} we discuss a high-level integration of AdapShare within the \gls{oran} architecture. Section~\ref{sec:conclusion} concludes the paper.
\section{Related Work}
\label{sec:prior_work}
Effective spectrum management is crucial for optimizing \gls{lte} and 5G \gls{nr} networks~\cite{ref:gopal2022}. Despite significant strides in \gls{lte}-\gls{nr} coexistence~\cite{ref:Coex_Xu_2021,ref:Coex_An_2020,ref:Coex_li_2021,ref:Coex_Ratasuk_2020}, several aspects can be improved, especially with the integration of \gls{oran} principles. Spectrum sharing within the \gls{oran} framework has recently garnered attention from the \gls{oran} Alliance Working Group~1 (WG1)~\cite{ref:oran_tr_wg1}, which identifies \gls{dss} as a key use case. Also, recent studies~\cite{ref:oran_smith_2021,ref:oran_mungari_2021_rl,ref:oran_baldesi_2022_charm,ref:oran_bonati_2023_neutran,ref:oran_kulacz_2022_dynamic,ref:oran_gopal_2024prosas} have delved into spectrum sharing with \gls{oran}. In~\cite{ref:oran_smith_2021}, the authors propose spectrum sharing between government \gls{LEO} satellites and the 5G \gls{ul}, employing O-RAN's intelligence and open interfaces. Frameworks compliant with O-RAN, such as those introduced in~\cite{ref:oran_baldesi_2022_charm}, demonstrate their efficacy in scenarios like LTE-WiFi coexistence in unlicensed frequency bands. The authors demonstrated the framework's performance using srsRAN for prototype development and the Colosseum channel emulator for data collection. Furthermore,~\cite{ref:oran_kulacz_2022_dynamic} explores contextual user density data for spectrum management in O-RAN-based networks, providing numerical validation of the proposed concept. In our previous work~\cite{ref:oran_gopal_2024prosas}, we proposed an intelligent radio resource demand
prediction and management scheme for intent-driven spectrum
management. The solution is a quasi-static scheme that often results in over-provisioning or under-provisioning of resources. To address these limitations, in this work, we employed an \gls{rl}-based approach, which dynamically allocates resources based on the predicted demand. Using a dynamic method allows for more efficient resource utilization than the quasi-static approach.

In addition to \gls{oran}, \gls{rl}~\cite{ref:sutton2018reinforcement} has also emerged as a powerful tool for addressing \gls{dss}~\cite{ref:rl_challita2021deep,ref:rl_mosleh2020dynamic,ref:rl_dong2022dynamic}. In~\cite{ref:rl_challita2021deep}, authors propose a deep \gls{rl} algorithm based on Monte Carlo Tree Search (MCTS) for spectrum sharing between \gls{lte} and \gls{nr} where a controller decides on the resource split between the networks in every subframe while accounting for future network states. In~\cite{ref:rl_mosleh2020dynamic}, authors propose an \gls{rl} based Q-learning spectrum allocation algorithm for carrier selection that allows Wi-Fi and \gls{lte} to operate concurrently in the unlicensed band. Actor-critic \gls{rl} algorithms, which are particularly suited for continuous action spaces, have demonstrated effectiveness in optimizing resource allocation as well. The authors in~\cite{ref:rl_dong2022dynamic} study a spectrum sharing scenario where primary and secondary users employ deep neural networks with an actor-critic framework that allows the secondary users to predict and adapt to the spectrum usage of the primary users.  

Other authors have explored integrating \gls{rl} with \gls{oran} to optimize resource allocation; \cite{ref:oran_mungari_2021_rl} presents an RL-based radio resource management solution leveraging the O-RAN platform. Building on these advancements, AdapShare employs state-of-the-art actor-critic \gls{rl} algorithms in the \gls{oran} environment to optimize spectrum use, yielding significant improvements in resource utilization and network efficiency.

\section{AdapShare: The Proposed Solution}
\label{sec:proposed_framework}
\gls{oran}~\cite{ref:oran_polese_2023} represents a transformative approach to modernizing \gls{ran} systems. Unlike conventional \gls{ran}s, which often depend on proprietary hardware and software solutions from a single vendor, \gls{oran} promotes the use of open interfaces and virtualization technologies. It introduces near-RT \gls{ric}s and non-RT \gls{ric}s to coordinate and manage network resources. These \gls{ric}s are pivotal in the \gls{oran} architecture, optimizing radio resource allocation, enhancing network performance, and ensuring \gls{qos}. Additionally, \gls{oran} fosters innovation by enabling the development of rApps and xApps that seamlessly interface with the \gls{ric}s and \gls{ran} infrastructure, providing advanced functionalities, intelligence, and adaptability. AdapShare is designed to leverage rApps, xApps, and open interfaces, thereby fully harnessing the capabilities of the O-RAN architecture.

In this paper, we propose AdapShare, an \gls{rl}-based intelligent radio resource allocation scheme designed for intent-driven (priority-based) spectrum management. It utilizes \gls{rl} agents to learn radio resource demand patterns and allocate resources to networks. We demonstrate its effectiveness in the context of spectrum sharing between \gls{lte} and \gls{nr} networks.
AdapShare employs a high-level model for \gls{lte} and \gls{nr} that makes few assumptions about the underlying technology. We assume that \gls{nr} employs \gls{lte}-compatible numerology with 15~kHz subcarrier spacing, resulting in identical time/frequency resource grids for both networks. The networks share a pool of time-frequency resources, referred to as \gls{prb}. We define resource demand as the number of \gls{prb} required by a network for data transmission. The \gls{rl} agent learns the resource usage patterns of each network by analyzing their individual resource usage time-series datasets and then allocates resources to minimize any surplus or deficit in \gls{prb} experienced by both networks. A surplus indicates resource over-provisioning, while a deficit represents under-provisioning, leading to packet loss or buffering. In the following sections, we formulate the resource allocation problem and discuss the \gls{rl}-based approach we used to solve it.

\subsection{Optimal Resource Allocation}
We describe how resources from a shared pool of \gls{prb} are distributed between coexisting \gls{lte} and \gls{nr} networks. We assume that the agent allocates \gls{prb} to meet each network's resource demands at regular time intervals. These intervals can represent time slots in an \gls{lte} system, $1$~ms subframes, periods of several seconds, minutes, or hours, depending on the scenario. Let $N_{r,t}$ represent the size of the \gls{prb} resource pool used by the agent to distribute resources ($N_{A,t}$ for \gls{lte} and $N_{B,t}$ for \gls{nr}) at time interval $t$. Our primary goal is to minimize the resource surplus or deficit for both networks by determining the optimal allocation using a performance metric that addresses the networks' varying demands. The optimization problem (OPT) is defined as follows:
\begin{eqnarray}
\textbf{OPT:}&\min_{N_{A,t}, N_{B,t}} J(N_{A,t}, N_{B,t} | \zeta, D_{A,t}, D_{B,t}),\nonumber\\
\textbf{s.t.}&\quad N_{A,t} + N_{B,t} \leq N_{r,t},\nonumber\\
&\quad 0 \leq N_{A,t} \leq N_{r,t}, \nonumber\\
&\quad 0 \leq N_{B,t} \leq N_{r,t},
\label{eqn:opt_prob}   
\end{eqnarray}
where 
\begin{equation}
J= \zeta\left(\frac{N_{A,t}-D_{A,t}}{D_{A,t}}\right)^2 
+ (1-\zeta) \left(\frac{N_{B,t}-D_{B,t}}{D_{B,t}}\right)^2,
    \label{eqn:obj_func_gen}
\end{equation}
is the weighted sum of squared fractional surpluses/deficits, i.e., $(N_{A,t}-D_{A,t})/D_{A,t}$ and $(N_{B,t}-D_{B,t})/D_{B,t}$, experienced by \gls{lte} and \gls{nr}, respectively. $(N_{A,t}, N_{B,t})$ is the ordered pair of allocation and $(D_{A,t},D_{B,t})$ is the ordered pair containing \gls{lte}'s and \gls{nr}'s respective demands at time $t$. $\zeta \in [0,1]$ is a weighting factor that enables intent-driven spectrum management by allowing the controller to prioritize one network over the other. Specifically, $\zeta \rightarrow 1$ assigns a higher priority to \gls{lte}, whereas, $\zeta \rightarrow 0$ assigns a higher priority to \gls{nr}. We solve the optimization problem~(\ref{eqn:opt_prob}) by using a learning-based approach.

\subsection{Reinforcement Learning Approach}
AdapShare's core objective is to optimize spectrum allocation by minimizing the surplus and deficit of radio resources between \gls{lte} and \gls{nr} networks. We formulate the optimization problem~(\ref{eqn:opt_prob}) as a contextual bandit problem, a variant of the multi-armed bandit problem that incorporates context information to guide decision-making~\cite{ref:sutton2018reinforcement}. The contextual bandit framework is well-suited to this scenario as it allows the \gls{rl} agents to make informed decisions (i.e., allocate resources) based on the environment's current and past states (the networks' resource demand). The agents' actions (resource allocation) do not affect future observations (networks' future resource demand), thus enabling more adaptive and efficient resource allocation. Next, we define the \gls{rl} setup. 

\textbf{State and observation:} In the contextual bandit formulation, the agent obtains knowledge about the state, i.e., the network resource demand, through observation. Here, the state space $S$ consists of ordered pairs of \gls{lte} and \gls{nr} resource demands. Each state $s_t \in S$ provides context, i.e., the networks' resource demand, that influences the decision-making process, i.e., resource allocation. At time $t$, the agent has observations of the networks' demand at $t$ and at $n$ previous time steps, given by $o_{t} = \{(D_{A,t},D_{B,t}),(D_{A,t-1},D_{B,t-1}),\dots,(D_{A,t-n},D_{B,t-n})\}$, where $(D_{A,k},D_{B,k})$ is the ordered pair containing \gls{lte}'s and \gls{nr}'s resource demand observed at time $k$ for $k = \{t,t-1,\dots,n\}$. 

\textbf{Action:} The variables of optimization in~(\ref{eqn:opt_prob}) are the resource allocation for \gls{lte} and \gls{nr}. Likewise, the action space $A$ consists of all possible partitions of the resource pool of $N_r$ \gls{prb} between the networks. Each action $a_t \in A$ corresponds to a specific allocation of $(N_{A,t},N_{B,t})$ \gls{prb} to the networks at time $t$, where the actions are continuous-valued. Depending on the value of $\zeta \in (0,1)$, $N_{A,t} \in (0,N_{r,t})$ and $N_{B,t} \in (0,N_{r,t})$, while satisfying the condition $N_{A,t} + N_{B,t} \le N_{r,t}$. At $\zeta = 0$, $N_{A,t} \ge 0$ and $N_{B,t} \le N_{r,t}$ and at $\zeta = 1$, $N_{A,t} \le N_{r,t}$ and $N_{B,t} \ge 0$. 

\textbf{Reward:} The reward function is crucial for \gls{rl} to make the learning algorithm a suitable replacement for the optimization problem~(\ref{eqn:opt_prob}). Since, in the contextual bandit formulation, neither rewards nor observations are influenced by the environment state (or by previous actions or observations), the environment does not evolve along the time dimension, and there is no sequential decision-making. Therefore, the problem focuses on finding the action that maximizes the current reward, i.e., finding $(N_{A,t},N_{B,t})$ that minimizes $J$ or maximizes $-J$. Hence, at time $t$, the reward is defined as: $r_t = - J - \eta J$,
where $\eta$ is the weighting factor that penalizes surplus/deficit. The reward reflects the efficiency of the resource allocation. It provides immediate feedback on the allocation's quality, guiding the \gls{rl} agents towards an optimal allocation that minimizes both surplus and deficit.

The goal of \gls{rl} is to find the optimal policy. By following this policy, the agent will take the best action when observing the system state. By following the ``state, action, reward, next-state'' trajectory, the optimal policy will maximize the cumulative discounted reward, which is given by: $R_t = \sum_{t=0}^{T} \gamma^{t} r_{t},\quad 0 \le \gamma \le 1,$
where $\gamma$ is the discount factor and $T$ is the time horizon.

\section{Methodology}
\label{sec:methodology}
This section outlines the methodology we used to develop and evaluate AdapShare, focusing on the design of the \gls{rl} agents, the data collection process, and the training and evaluation procedures.

\subsection{RL Agent Design}
AdapShare utilizes actor-critic \gls{rl} methods~\cite{ref:sutton2018reinforcement} that combine aspects of both policy-based methods (Actor) and value-based methods (Critic). In the actor-critic framework, the \gls{rl} agent or the ``actor'' learns a policy to make decisions, and a value function or the ``critic'' evaluates the actions taken by the actor. Simultaneously, the critic evaluates these actions by estimating their value or quality. This dual role allows the agent to strike a balance between exploration and exploitation, leveraging the strengths of both policy and value functions. The actor network, parameterized by $\theta$, selects actions based on the current policy $\pi_\theta(a_t | s_t)$. The critic network, parameterized by $\phi$, evaluates the selected actions by estimating the value function $V(s_t)$ that estimates the expected cumulative reward starting from state $s_{t}$. 

The Actor-Critic algorithm's objective function combines the policy gradient for the actor and the value function for the critic. For the actor, the objective function is expressed as the gradient of the expected return, calculated using the policy function and the advantage function, which reflects the benefit of taking specific actions in given states. This is mathematically represented by the policy gradient equation: $\nabla_{\theta} J(\theta) \approx \frac{1}{N} \sum_{i=0}^{N} \nabla_{\theta} \log \pi_{\theta}(a_i \mid s_i) \cdot A(s_i, a_i)$. For the critic, the objective is to minimize the loss between the estimated value and the actual action-value, expressed by $\nabla_{\phi} J(\phi) \approx \frac{1}{N} \sum_{i=1}^{N} \nabla_{\phi} (V_\phi(s_i) - Q_\phi(s_i, a_i))^2$. The update rules involve adjusting the actor's parameters using gradient ascent, $\theta_{t+1} = \theta_{t} + \alpha \nabla_{\theta} J(\theta_{t})$, and the critic's parameters using gradient descent, $\phi_{t} = \phi_{t} - \beta \nabla_{\phi} J(\phi_{t})$, where $\alpha$ and $\beta$ are the learning rates for the actor and critic, respectively. This combination allows the Actor-Critic algorithm to effectively learn and adapt policies for optimal decision-making.

Specifically, AdapShare employs two state-of-the-art actor-critic \gls{rl} agents: \gls{ddpg} and \gls{td3}~\cite{ref:rl_dong2022dynamic}. We chose these agents for their ability to handle continuous action spaces and their effectiveness in learning complex policies. \gls{ddpg}~\cite{ref:rl_dong2022dynamic} is an off-policy algorithm that leverages deep neural networks to approximate the policy and value functions. The core components of \gls{ddpg} include the actor network and the critic network. \gls{td3}~\cite{ref:rl_dong2022dynamic} enhances \gls{ddpg} by addressing its instability and overestimation issues. Key improvements include double Q-learning, delayed policy updates, and target policy smoothing. \gls{td3} employs two critic networks to mitigate the overestimation bias in value estimates, using the minimum value between the two critic networks to update the actor network. The policy update frequency is reduced to every two iterations, ensuring more stable training. Additionally, noise is added to the target action, reducing the variance of the target value and further stabilizing training.
\begin{figure*}
     \begin{subfigure}[b]{0.5\textwidth}
     \centering
         \includegraphics[scale=0.205]{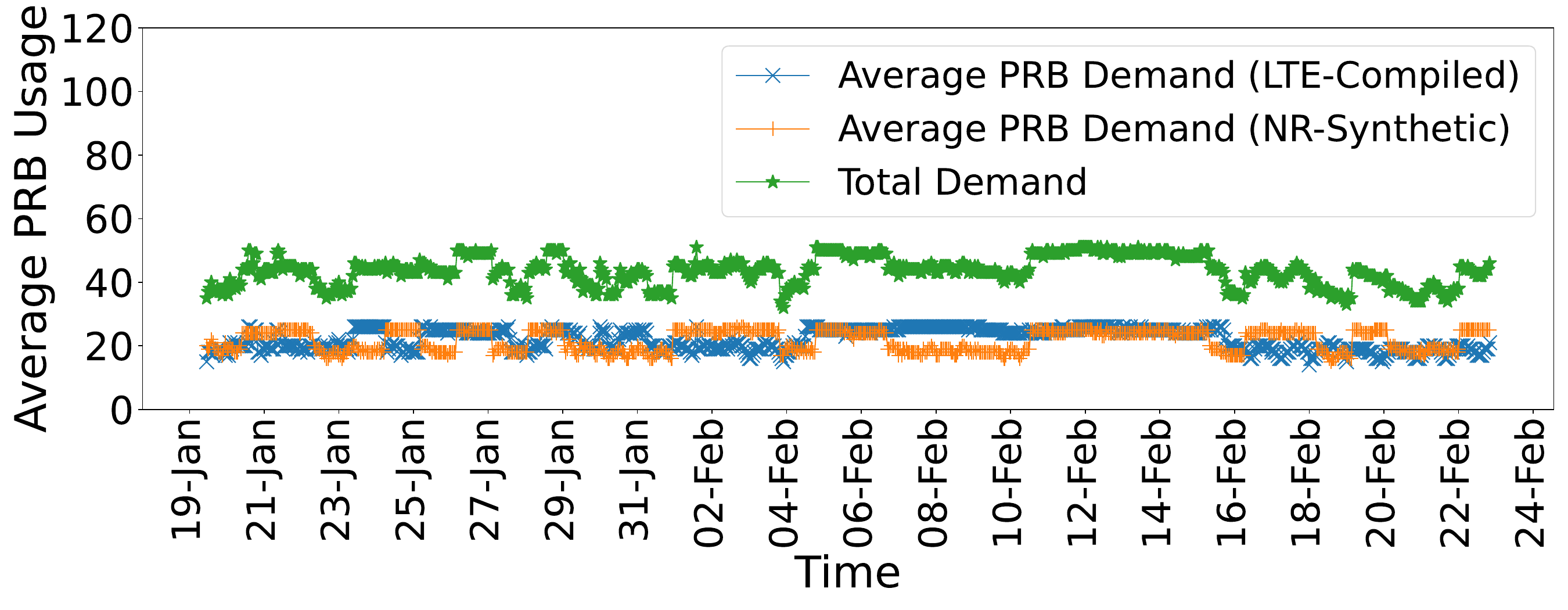}
         \caption{Average PRB usage vs. Time}
         \label{fig:lte_nr_demand}
     \end{subfigure}\hfill
     \begin{subfigure}[b]{0.5\textwidth}
     \centering
         \includegraphics[scale=0.21]{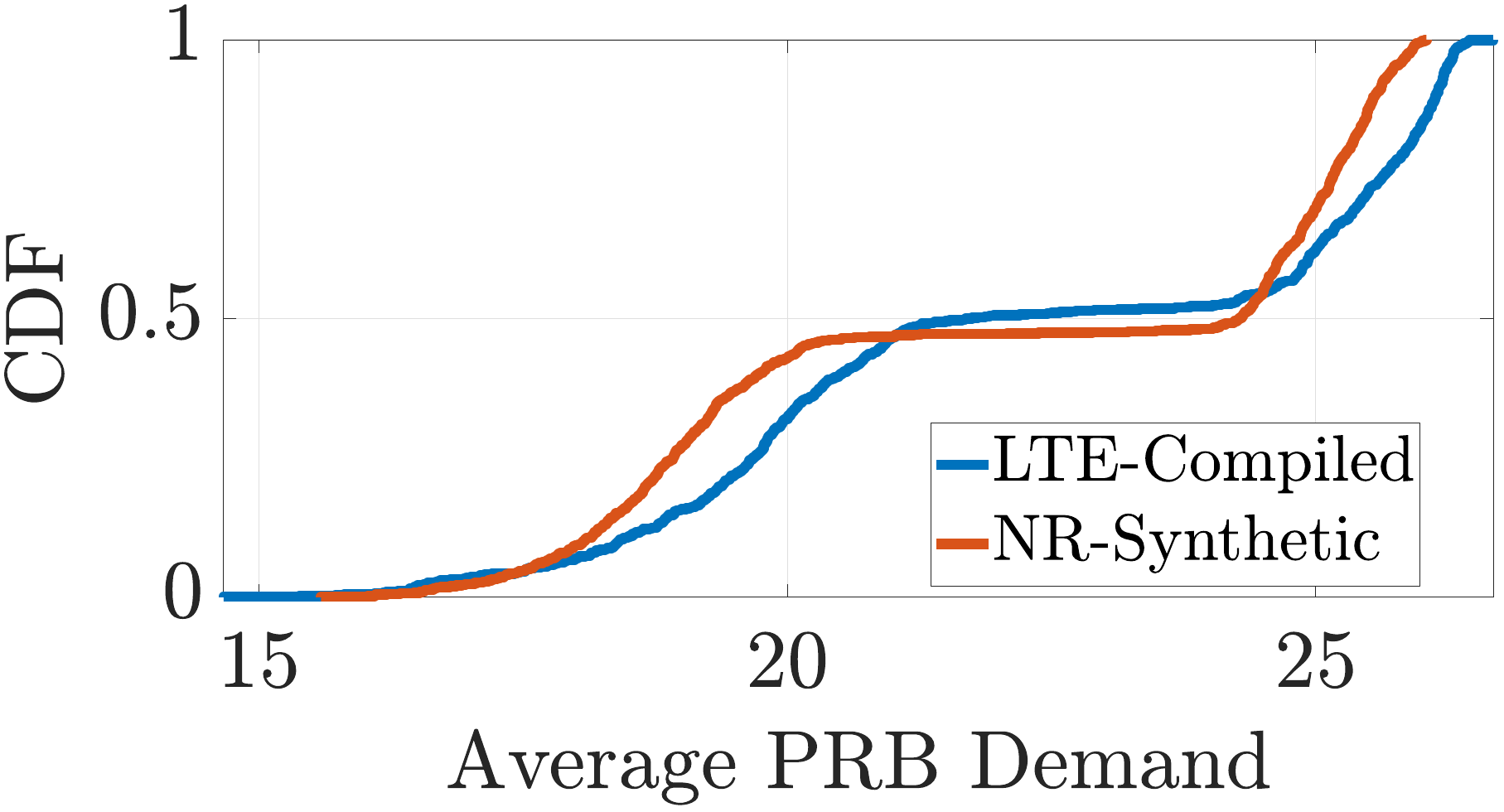}
         \caption{CDF of average PRB demand for \gls{lte} and \gls{nr} dataset.}
         \label{fig:cdf_lte_nr_demand}
     \end{subfigure}
        \caption{\small (a) Average PRB usage vs. time for \gls{lte}, \gls{nr} and \gls{lte}+\gls{nr}, i.e., aggregate, and (b) CDF of average PRB demand for \gls{lte} and synthetically generated \gls{nr} dataset. Both plots correspond to 1-hour time granularity.}
        \label{fig:Demand and CDF}
        \vspace{-1.25em}
\end{figure*}

\subsection{Data Collection}
For our study, we compiled an extensive dataset~\cite{ref:datset_adapshare} of \gls{lte} scheduling information that we collected at the NIST Gaithersburg campus between January and February 2023. This dataset is from downlink traffic at frequency 2115~MHz in Band~4 (2110–2155~MHz) and was collected using OWL~\cite{ref:owl_bui_2016}, an online decoder of the \gls{lte} control channel. We employed a BladeRF SDR board to transfer the collected real-world \gls{lte} signals to a laptop running Ubuntu~20.04 and used OWL to decode the signals. OWL captured the \gls{lte} \gls{dci} broadcast by the base station at the 1~ms subframe level. The dataset~\cite{ref:datset_adapshare} includes information such as the system frame number (SFN), subframe index, radio network temporary identifiers (RNTIs), the number of \gls{prb} allocated to devices, modulation and coding scheme (MCS), and DCI message type. We extracted the number of \gls{prb} corresponding to data transmissions, specifically \gls{dci} format~2B. We also included time stamps reflecting when this data was collected, resulting in a time series dataset detailing PRB allocations for the \gls{lte} network. Importantly, this data contains no user-specific information, ensuring complete anonymity.

Using the compiled real-world \gls{lte} dataset, we generated a synthetic PRB allocation dataset corresponding to the \gls{nr} network. To create this synthetic \gls{nr} dataset, we leveraged \gls{timegan}~\cite{ref:timegan}, a tool that generates synthetic time-series data by capturing temporal dependencies through \gls{rnn}. We used \gls{timegan} and the collected real-world LTE time-series data to produce synthetic time-series \gls{nr} data that closely mirrors the statistical properties and patterns observed in the \gls{lte} data. This approach is considered valid under the assumption that \gls{nr} employs \gls{lte}-compatible numerology with 15~kHz subcarrier spacing and shares an identical time-frequency resource grid. Additionally, it is assumed that \gls{nr} systems would serve devices similar to those in \gls{lte}.

For our analysis, we resampled the time-series data, transitioning from PRB allocation per millisecond to PRB allocation per hour by using the mean function. Consequently, we obtained a dataset representing the average PRB allocation for both \gls{lte} and \gls{nr} per hour. The dataset comprises approximately 860~samples. 
Fig.~\ref{fig:lte_nr_demand} displays the average PRB usage/demand over time for \gls{lte}, \gls{nr}, and the combined (\gls{lte}+\gls{nr}) datasets at 1-hour granularity. Additionally, Fig.~\ref{fig:cdf_lte_nr_demand} shows the \gls{cdf} of PRB demands from both networks. This graph highlights the similarity in demand distribution between the compiled \gls{lte} dataset and the synthetically generated \gls{nr} dataset, thereby confirming the effectiveness of \gls{timegan}. Next, we describe the training of the \gls{rl} agents and our performance evaluation of AdapShare.

\subsection{Training and Evaluation}
The training procedure combines supervised learning (for initial policy approximation) and reinforcement learning (for policy refinement). The \gls{rl} agents' training begins with the initialization of the actor and critic networks with random weights. An experience replay buffer is used to store past experiences, which allows for efficient learning by reusing previous experiences. Batch learning is employed, where mini-batches of experiences are sampled from the replay buffer to update the networks. This approach ensures stable and efficient learning by preventing overfitting and ensuring that the networks generalize well to new states~\cite{ref:datset_adapshare}.

To evaluate AdapShare's performance, we compute the average surplus/deficit, denoted by $S_{A}$ and $S_{B}$ for \gls{lte} and \gls{nr}, respectively, which measures the amount of over-provisioned or under-provisioned resources, and defined as
\begin{equation}
\small
S_{A} = \frac{1}{T}\sum\limits_{t=1}^{T}\left(\frac{N_{A,t}^{*}-D_{A,t}}{D_{A,t}}\right),
S_{B} = \frac{1}{T}\sum\limits_{t=1}^{T}\left(\frac{N_{B,t}^{*}-D_{B,t}}{D_{B,t}}\right)\,.
\label{eq:surplus_deficit}
\normalsize
\end{equation}
where, $(N_{A,t}^*,N_{B,t}^*)$ denote the optimal allocation by the \gls{rl} agents. 

Additionally, we use Jain's Fairness Index to measure the fairness of the resource allocation between \gls{lte} and \gls{nr} networks:
\begin{equation}
    F = \frac{1}{T}\sum\limits_{t=1}^{T}\left(\frac{(N_{A,t}^* + N_{B,t}^*)^2}{2\bigl((N_{A,t}^*)^2 + (N_{B,t}^*)^2\bigr)}\right),\quad \forall \zeta \in [0,1].
    \label{eq:fairness}
\end{equation}
This metric ensures that for every value of $\zeta$, the allocation is balanced between the two networks.

Finally, we compare AdapShare's performance with a quasi-static resource allocation scheme based on network demand statistics~\cite{ref:oran_gopal_2024prosas}, i.e., the maximum demand. The optimization problem, $\text{OPT}_{\text{base}}$, represents the baseline and can be obtained by replacing $(D_{A,t},D_{B,t})$ with $(M_{D_{A}},M_{D_{B}})$ in~(\ref{eqn:opt_prob}) and~(\ref{eqn:obj_func_gen}), where the latter corresponds to the maximum resource demand for \gls{lte} and \gls{nr}, respectively. This baseline allocates resources based on historical demand patterns without adapting to real-time changes, which provides a benchmark for evaluating the benefits of AdapShare's adaptive approach in terms of efficiency and fairness.
\section{Results}
\label{sec:results}
\begin{figure*}
     \begin{subfigure}[b]{0.32\textwidth}
     \centering
         \includegraphics[scale=0.18]{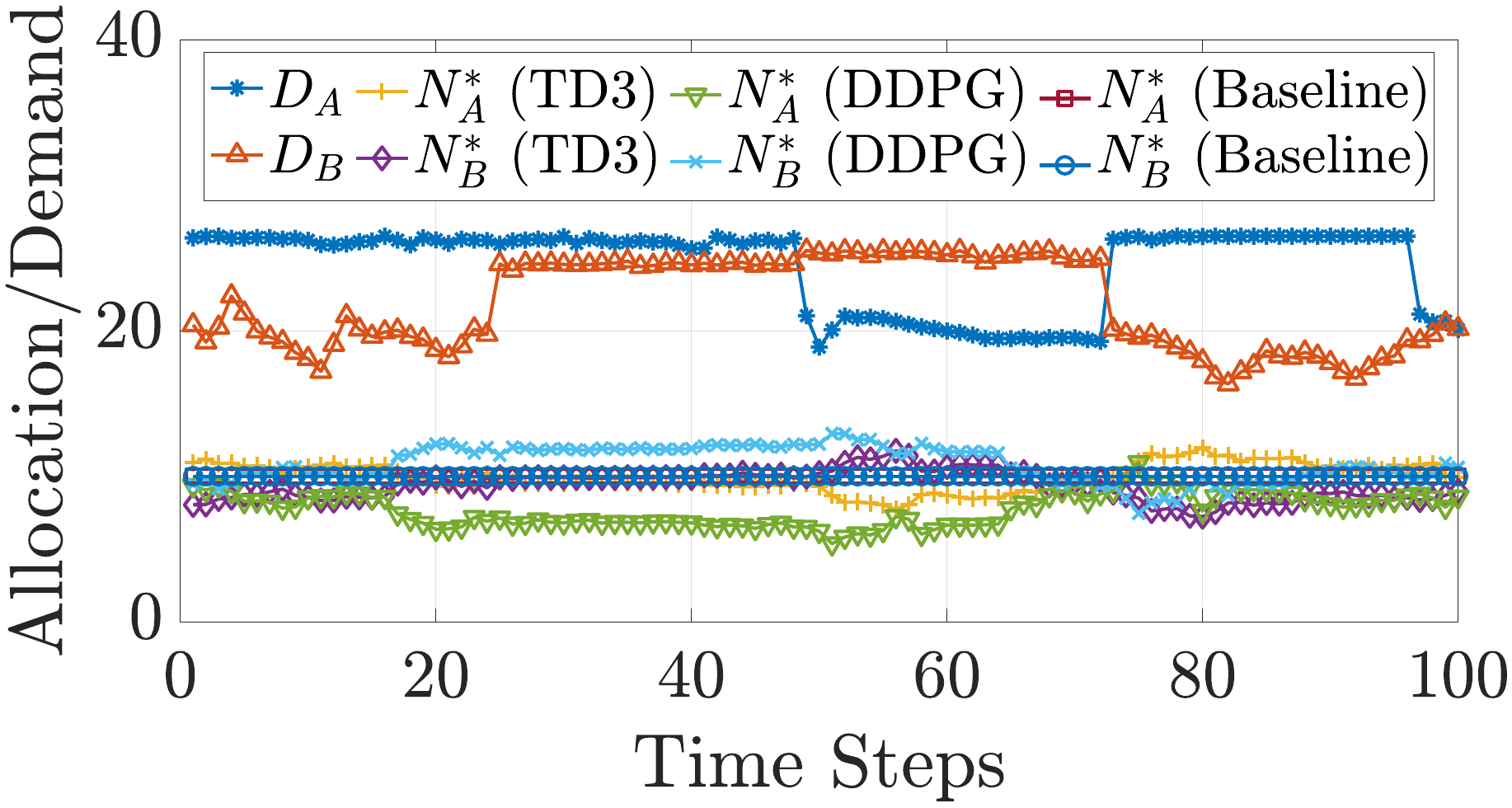}
         \caption{\small Resource demand and allocation for $N_r = 20$ and $\zeta = 0.5$.}
         \label{fig:alloc_20}
     \end{subfigure}
     \enspace
     \begin{subfigure}[b]{0.32\textwidth}
     \centering
         \includegraphics[scale=0.18]{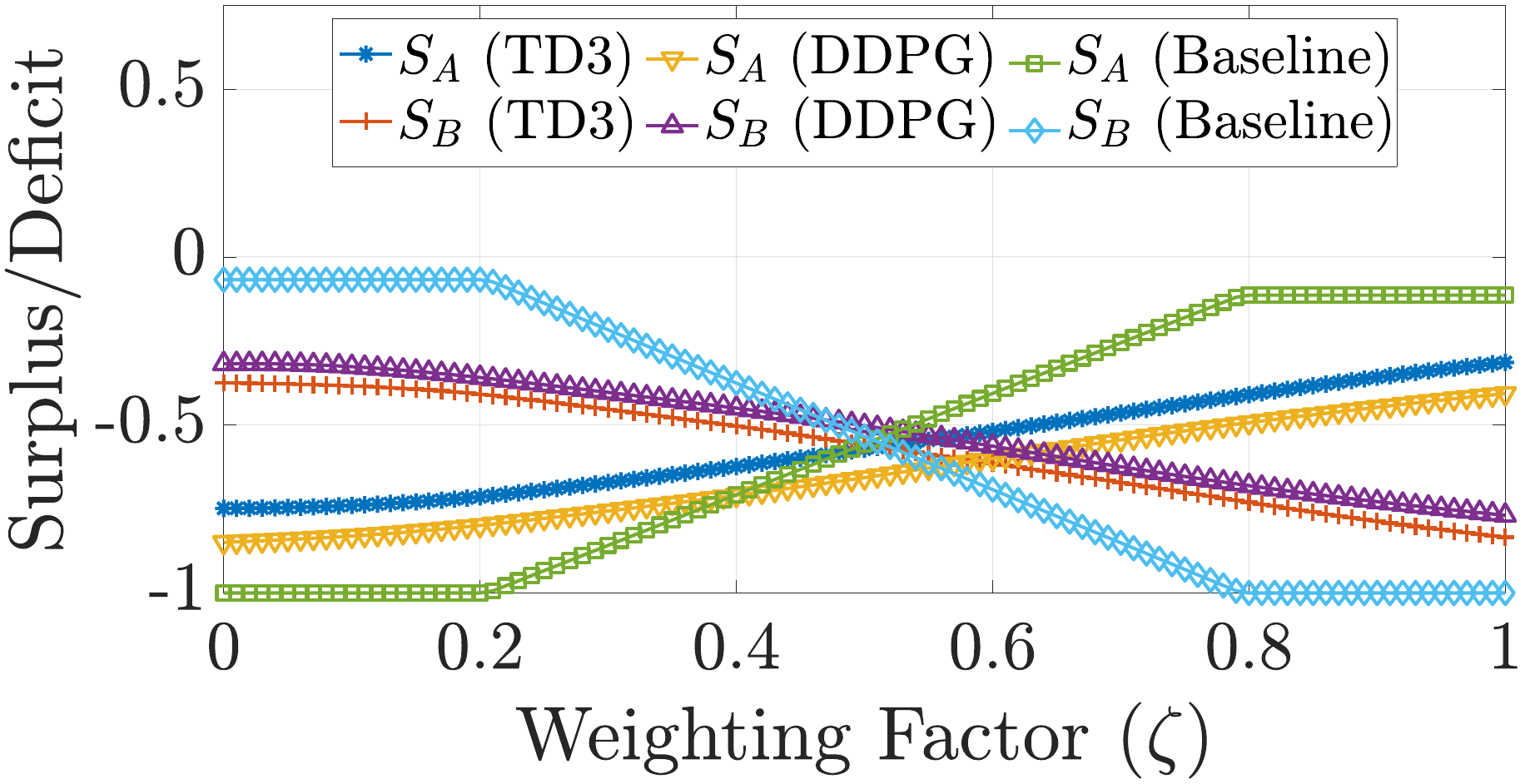}
         \caption{\small Surplus/Deficit vs. weighting factor $\zeta$ for $N_r = 20$.}
         \label{fig:sur_def_20}
     \end{subfigure}
     \enspace
     \begin{subfigure}[b]{0.32\textwidth}
     \centering
         \includegraphics[scale=0.18]{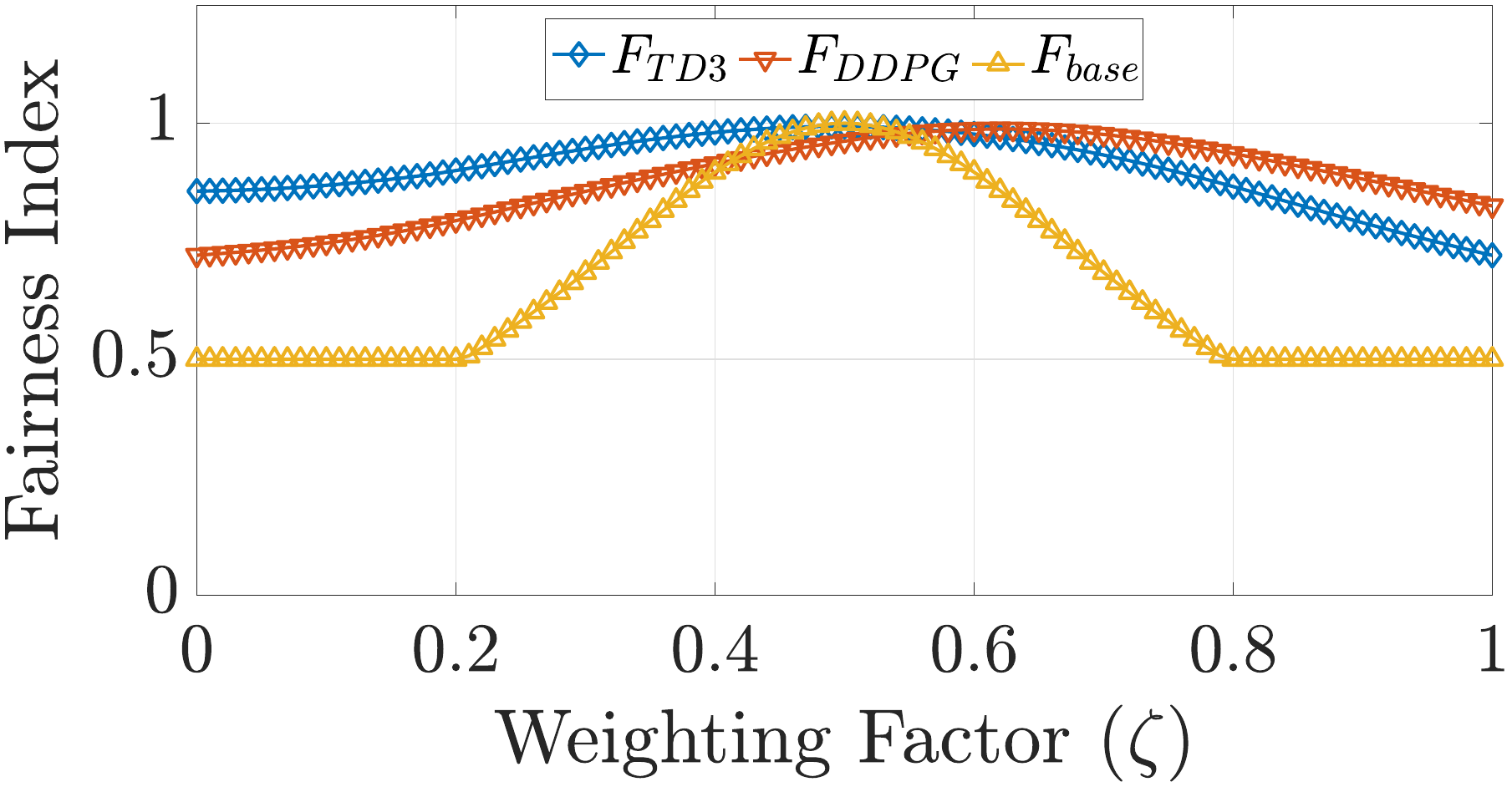}
         \caption{\small Jain's Fairness index vs. weighting factor $\zeta$ for $N_r = 20$.}
         \label{fig:fair_20}
     \end{subfigure}\\
     \begin{subfigure}[b]{0.32\textwidth}
     \centering
         \includegraphics[scale=0.18]{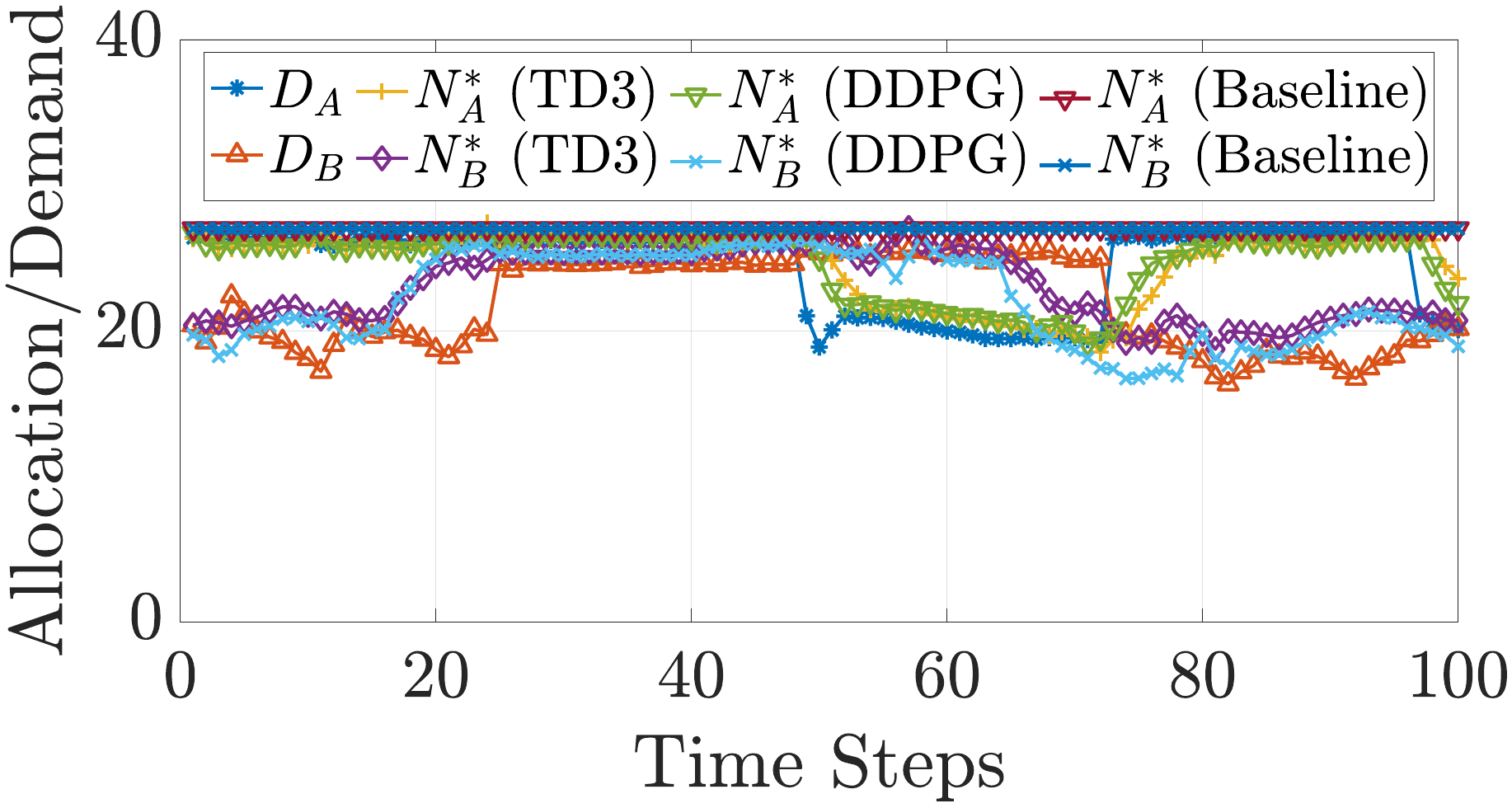}
         \caption{\small Resource Demand and allocation for $N_r = 100$ and $\zeta = 0.5$.}
         \label{fig:alloc_100}
     \end{subfigure}
     \enspace
     \begin{subfigure}[b]{0.32\textwidth}
     \centering
         \includegraphics[scale=0.18]{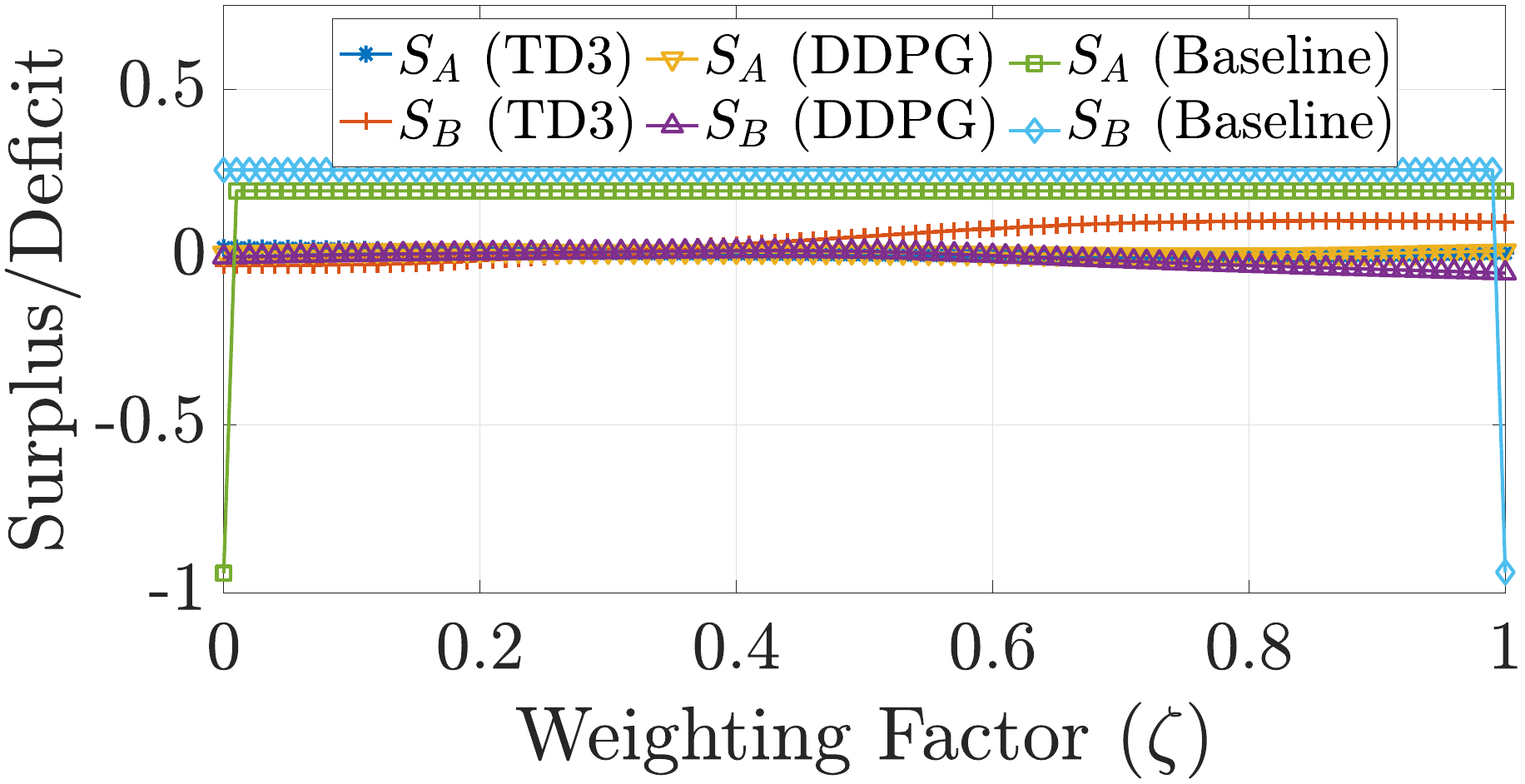}
         \caption{\small Surplus/Deficit vs. weighting factor $\zeta$ for $N_r = 100$.}
         \label{fig:sur_def_100}
     \end{subfigure}
     \enspace
     \begin{subfigure}[b]{0.32\textwidth}
     \centering
         \includegraphics[scale=0.18]{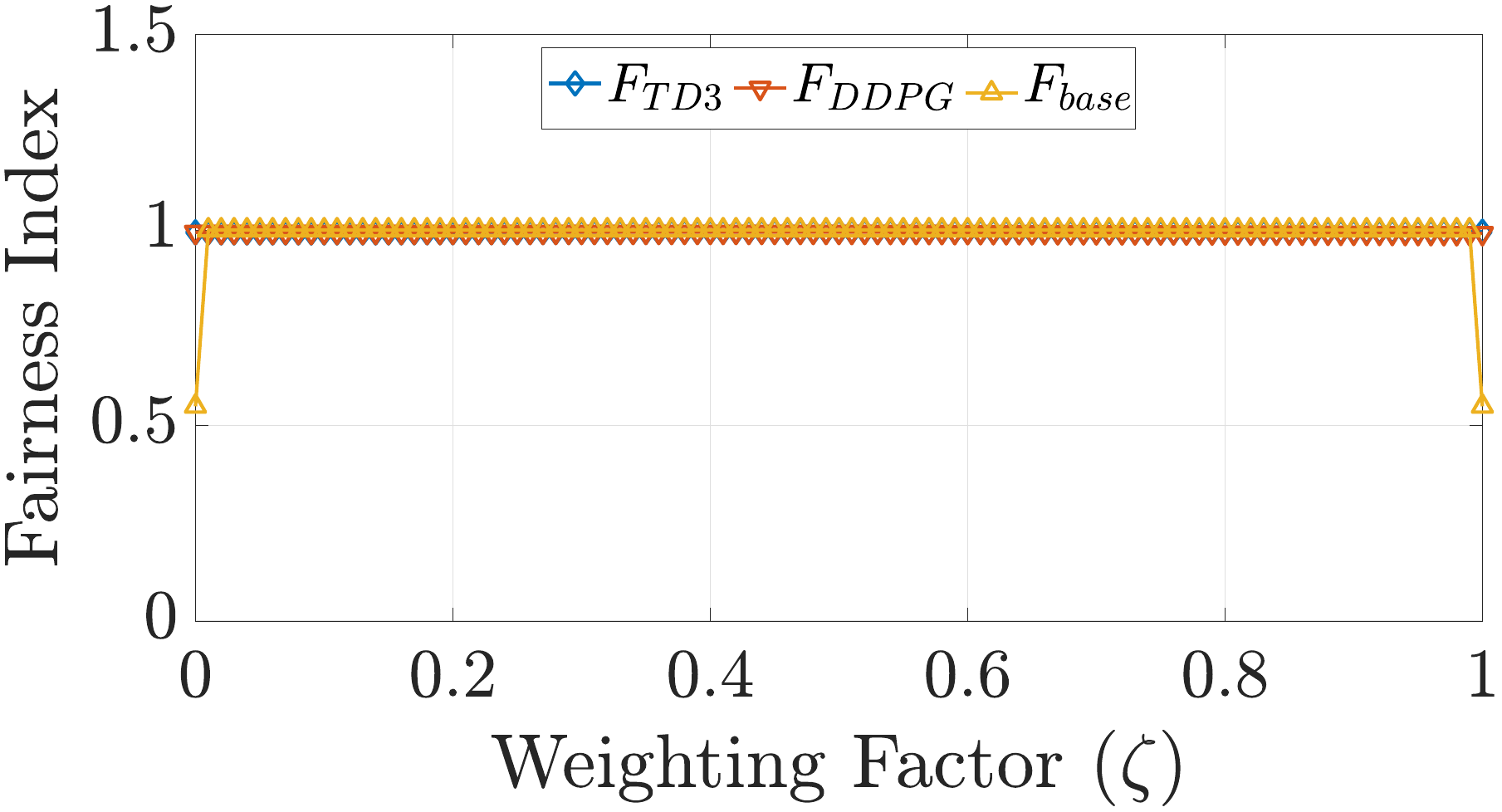}
         \caption{\small Jain's Fairness index vs. weighting factor $\zeta$ for $N_r = 100$.}
         \label{fig:fair_100}
     \end{subfigure}
        \caption{\small Resource demand and allocation, surplus/deficit, and Jain's Fairness Index for $N_r = \{20,100\}$.}
        \label{fig:results}
        \vspace{-1.25em}
\end{figure*}

We assessed the effectiveness of AdapShare by conducting a comprehensive analysis across three distinct resource pool sizes, $N_r \in \{20, 60, 100\}$ \gls{prb}, representing limited, moderate, and adequate resource availability. For each pool size, we varied the parameter $\zeta$ from $0$ to $1$.  For each combination of $N_r$ and $\zeta$ values, we trained the \gls{rl} agents, \gls{ddpg} and \gls{td3}, using real-world \gls{lte} data and synthetic \gls{nr} datasets. Once trained, we employed the agents to obtain the optimal resource partitions, $N_A^*$ and $N_B^*$ resources. Our evaluation aimed to capture the average surplus or deficit of resource allocation~(\ref{eq:surplus_deficit}) and Jain's fairness index~(\ref{eq:fairness}) for different $\zeta$ values and resource pool sizes. For brevity, we present results corresponding to resource demand and allocation for \gls{lte} and \gls{nr} with $N_r = \{20, 100\}$ \gls{prb} and $\zeta = 0.5$, along with the surplus/deficit and Jain's fairness index obtained using both \gls{rl} agents, as depicted in Fig.~\ref{fig:results}. Key parameters characterizing the demand profiles of \gls{lte} and \gls{nr} include a maximum demand, $M_{D_A} = 26.31$ and $M_{D_B} = 25.80$~resources, with an expected total demand of approximately $45$ resources per hour. Note that while we consider the dataset representing the average \gls{prb} allocation for both \gls{lte} and \gls{nr} per hour, our inferences are applicable to other time granularities, such as per minute and second.

\begin{figure}[t]
    \centering
    \includegraphics[width=\columnwidth]{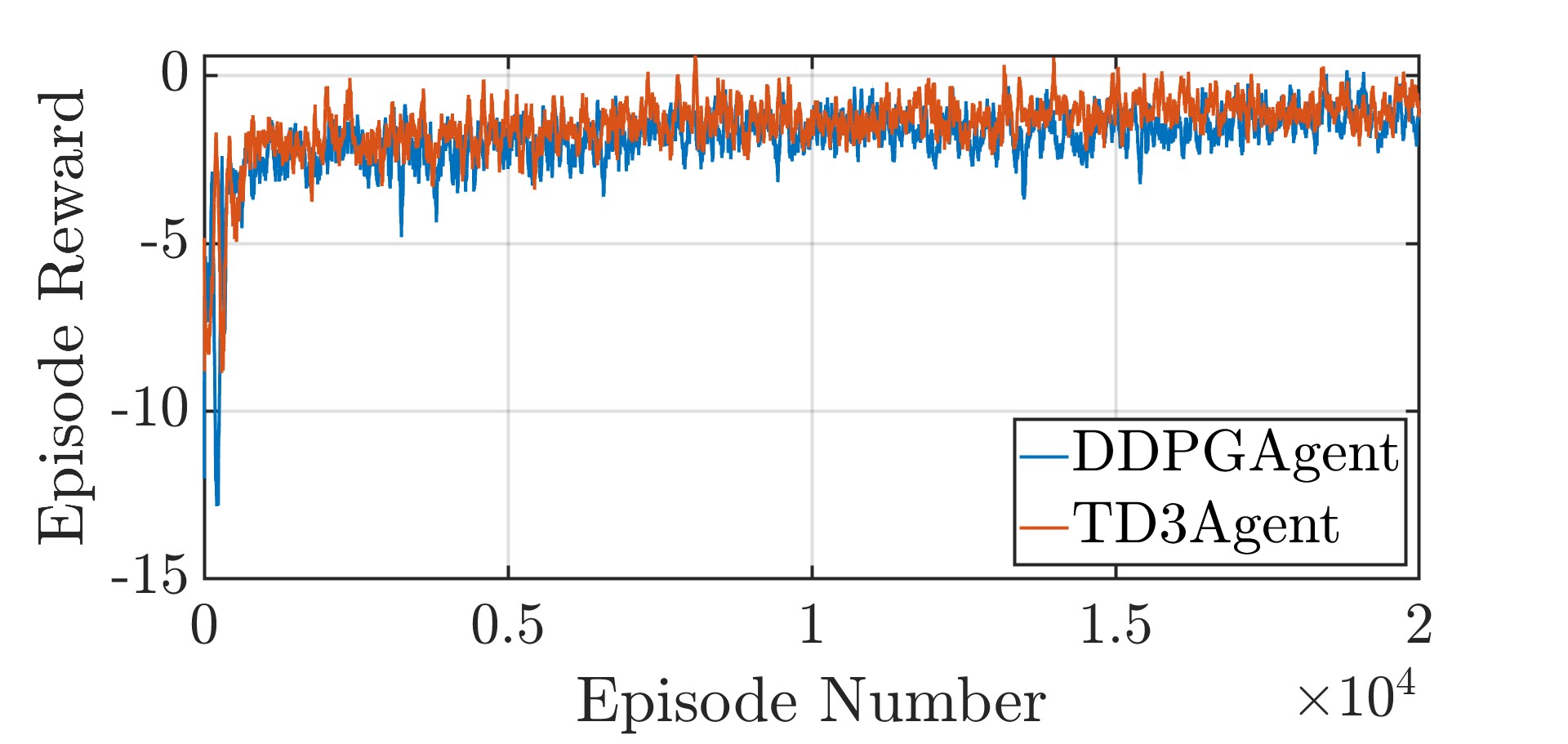}
    \caption{\small Performance Comparison between \gls{ddpg} and \gls{td3}.}
    \label{fig:rl_agent_comparison}
\vspace{-1.25em}
\end{figure}
In scenarios with a limited resource pool ($N_r = 20$), where resources are scarce and often insufficient to meet the demand, both networks experience starvation. Total starvation, which occurs when the networks receive no resources, corresponds to a fractional deficit of~$-1$. Our findings (Fig.~\ref{fig:alloc_20}) indicate that with $\zeta = 0.5$, the \gls{rl} agents' allocation aligns more closely with the network demands compared to the baseline ($\text{OPT}_{\text{base}}$), which splits resources equally. The \gls{rl} agents prevent total starvation across all $\zeta$ values (Fig.~\ref{fig:sur_def_20}), whereas $\text{OPT}_{\text{base}}$ frequently results in total starvation. Moreover, the \gls{rl} agents achieve a higher fairness index (Fig.~\ref{fig:fair_20}). For $N_r = 60$~resources, where the resource pool is moderate in size and can accommodate the maximum demands, the resource allocation by the \gls{rl} agents aligns closely with the network demand, leading to a slight deficit. In contrast, $\text{OPT}_{\text{base}}$ allocates resources equal to the maximum demand, resulting in a surplus. Lastly, in scenarios with adequate resources ($N_r = 100$), the resource pool can meet the maximum demands of both networks, and the \gls{rl} agents effectively matched the allocation to the network demands (Fig.~\ref{fig:alloc_100}), preventing the over-provisioning that we observed with $\text{OPT}_{\text{base}}$ (Fig.~\ref{fig:sur_def_100}). Using Jain's fairness index~(\ref{eq:fairness}), with higher values indicating more equitable distribution, both \gls{rl} agents outperformed $\text{OPT}_{\text{base}}$, particularly in scenarios with limited resources and adequate resources, achieving higher fairness indices across different $\zeta$ values (Fig.~\ref{fig:fair_20} and~\ref{fig:fair_100}). The \gls{td3} agent consistently achieved the highest fairness scores, demonstrating its effectiveness in balancing resource allocation between \gls{lte} and \gls{nr} networks.

Fig.~\ref{fig:rl_agent_comparison} presents a performance comparison in terms of the learning curve between the \gls{ddpg} and \gls{td3} agents. The analysis revealed that both agents effectively minimized resource surplus and deficit while optimizing for fairness. However, \gls{td3} slightly outperformed \gls{ddpg} in terms of stability and achieving higher fairness indices across various $\zeta$ values. AdapShare's adaptive approach outperforms the quasi-static baseline ($\text{OPT}_{\text{base}}$) in terms of both efficiency and fairness. The \gls{rl} agents effectively minimize resource surplus and deficit, demonstrating the benefits of dynamic allocation, particularly when resources are limited or adequate. These results highlight the potential of AdapShare in enhancing resource allocation efficiency and fairness in \gls{oran} environments.

\section{Integration with \gls{oran} Architecture}
\label{sec:integration}
AdapShare is designed to seamlessly integrate with the \gls{oran} architecture, leveraging its open interfaces, virtualization, and AI/ML capabilities to enhance spectrum sharing between \gls{lte} and \gls{nr} networks. This integration is pivotal to achieving dynamic and intelligent spectrum management, ensuring optimal utilization of radio resources in near real-time. AdapShare can be deployed flexibly within this architecture, either as a rApp in the non-RT \gls{ric} or as an xApp in the near-RT \gls{ric}.

\begin{figure*}[t]
    \centering
    \includegraphics[width=0.9\textwidth]{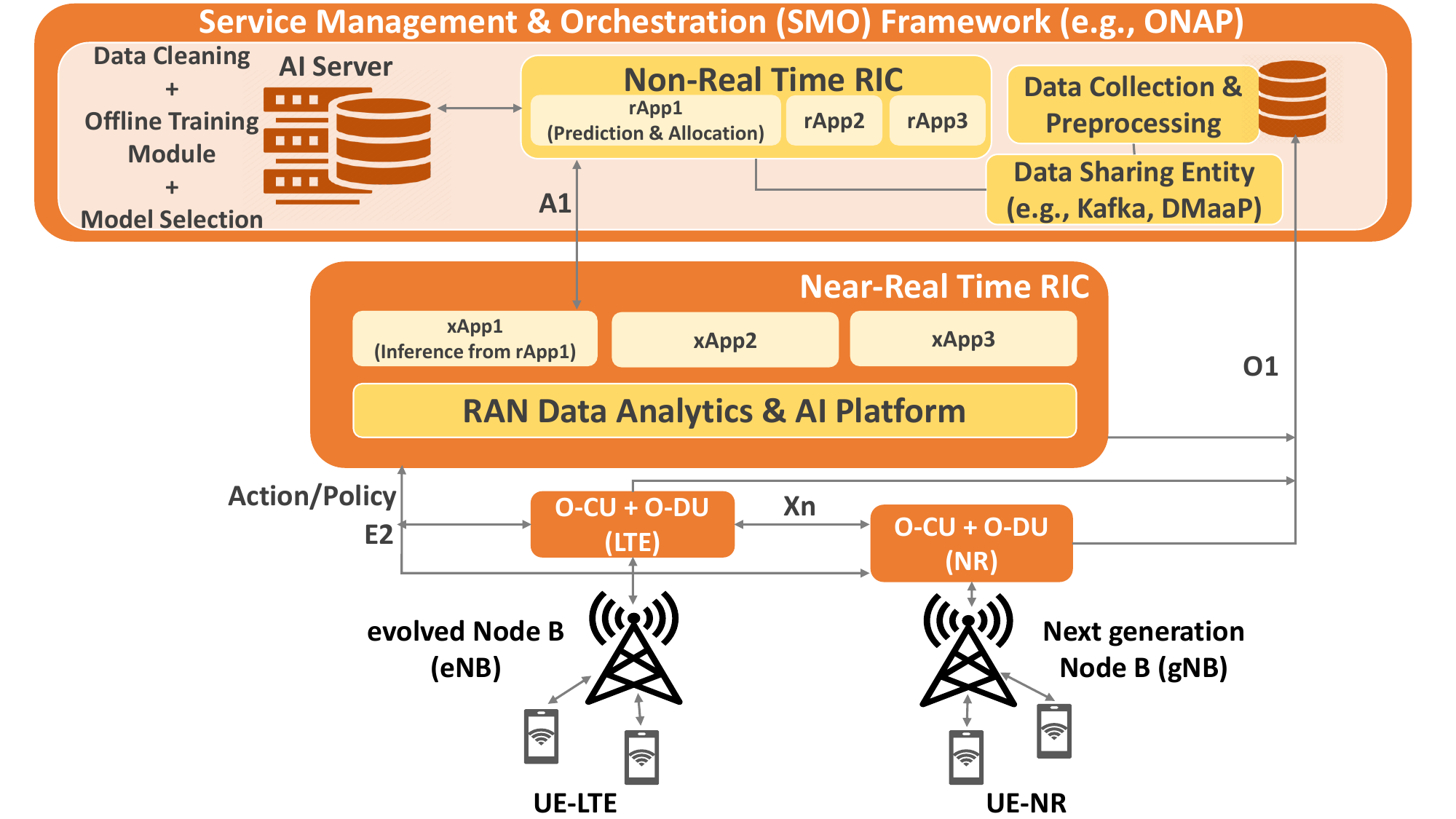}
    \caption{\small High-level structure illustrating the deployment of AdapShare as a rApp within the \gls{oran} architecture~\cite{ref:oran_tr_wg1}.}
    \label{fig:deployment}
\vspace{-1.25em}
\end{figure*}

In the non-RT RIC deployment scenario, AdapShare functions as a rApp 
that hosts the \gls{rl} agent that learns and analyzes the network state over extended periods, generating insights and optimization policies. These policies are communicated to the near-RT \gls{ric} for implementation. Initially, data collected from \gls{ran} components (the O-RAN Central Unit (O-CU) and O-RAN Distributed Unit (O-DU)), including the \gls{dci}, is transmitted to the data collector in the \gls{smo} entity through the O1 interface. This data is then routed to the non-RT \gls{ric} in the \gls{smo} via a data-sharing entity such as \gls{dmaap} or Kafka. The non-RT \gls{ric} queries the \gls{rl} agent in the AI server (e.g., AcumosAI) within the \gls{smo}, and once the agent is trained, the non-RT \gls{ric} is notified of the inference. The inference results and associated policies are sent to the Resource Allocation xApp in the near-RT \gls{ric} via the A1 interface. The Resource Allocation xApp communicates the allocation to the RAN components using the E2 interface. Fig.~\ref{fig:deployment} illustrates a high-level deployment of AdapShare within the \gls{oran} architecture as a rApp. The design and deployment are based on the guidelines outlined in the TR~\cite{ref:oran_tr_wg2} by the \gls{oran} Alliance. This deployment allows for extensive data analysis and the development of sophisticated optimization strategies without stringent latency constraints. 

Alternatively, AdapShare can be deployed as an xApp in the near-RT \gls{ric}. Here, the \gls{rl} agents operate in a near-RT environment, observing network conditions and making immediate adjustments to resource allocation. This deployment scenario ensures rapid response to changing network demands, which is crucial for maintaining optimal performance in dynamic environments.

\section{Conclusions}
\label{sec:conclusion}
In this paper, we introduced AdapShare, a novel radio resource allocation scheme tailored for intent-driven spectrum management within the \gls{oran} architecture. By leveraging state-of-the-art actor-critic \gls{rl} algorithms, namely \gls{ddpg} and \gls{td3}, AdapShare dynamically learns and adapts to varying network demands, optimizing spectrum allocation to minimize resource surpluses and deficits. Our comprehensive analysis demonstrated that AdapShare's \gls{rl} agents significantly enhance resource allocation efficiency and fairness compared to a quasi-static baseline, effectively reducing resource starvation and over-provisioning. AdapShare also achieved greater fairness in resource distribution, which is crucial for maintaining equitable service levels in heterogeneous network environments. By integrating AdapShare into the \gls{oran} architecture, we highlighted its practical utility and scalability for real-world deployments, making it a promising solution for enhancing spectrum management in future 5G and beyond networks. This work underscores the promising integration of \gls{rl} techniques with \gls{oran} and paves the way for future advancements in dynamic spectrum management.
\section*{Acknowledgement}
This work is partially funded by the Department of Homeland Security’s Science and Technology Directorate (S\&T).
\begin{spacing}{}
\bibliographystyle{IEEEtran}
\bibliography{references}
\end{spacing}
\end{document}